\documentclass[dvips]{arxstspdf}
\usepackage{stfloats,flushend}
\usepackage{graphicx}

\volume{25}
\issue{1}
\pubyear{2010}
\firstpage{107}
\lastpage{125}
\doi{10.1214/10-STS326}

\makeatletter
\newtheorem{lemma}{Lemma}[section]

\newcommand{\niekcite}[1]{\citeauthor{#1}, \citeyear{#1}}
\renewcommand{\citep}[1]{(\citeauthor{#1}, \citeyear{#1})}

\newcommand{\bld}[1]{\mathbf{#1}}

\newcommand{\smttt}[1]{\mathrm{#1}}
\newcommand{\N}{\mathcal{N}}
\newcommand{\matern}{Mat\'{e}rn }

\newcommand{\bauthore}[1]{\textsc{#1}}
\newcommand{\bsnme}[1]{#1}
\newcommand{\bfnme}[1]{}
\newcommand{\binitse}[1]{#1}
\newcommand{\byeare}[1]{#1}
\newcommand{\bjournale}[1]{\textit{#1}}
\newcommand{\bvolumee}[1]{\textbf{#1}}
\newcommand{\bpagese}[1]{#1}

\newcommand{\ANDe}{\textup{and} }
\newcommand{\btitlee}[1]{#1}
\newcommand{\beditione}[1]{#1}

\newcommand{\bpublishere}[1]{#1}
\newcommand{\baddresse}[1]{#1}
\newcommand{\btypee}[1]{#1}
\newcommand{\bnumbere}[1]{#1}
\newcommand{\binstitutione}[1]{#1}
\newcommand{\betal}[1]{\textsc{#1}}

\newcommand{\bbooktitlee}[1]{\textit{#1}}
\newcommand{\beditor}[1]{#1}

\makeatother

\begin{document}
\begin{frontmatter}

\title{The Importance of Scale for Spatial-Confounding Bias and Precision\break
of Spatial Regression Estimators}
\runtitle{Scale and spatial regression estimators}

\begin{aug}
\author{\fnms{Christopher J.} \snm{Paciorek}\corref{}\ead[label=e1]{paciorek@alumni.cmu.edu}}
\runauthor{C. J. Paciorek}

\affiliation{Harvard School of Public Health and University of
California, Berkeley}

\address{Christopher J. Paciorek is Visiting Assistant Professor,
Department of Biostatistics, Harvard School of Public Health,
655 Huntington Avenue, Boston, Massachusetts 02115, USA and
Adjunct Assistant Professor, Department of Statistics,
367 Evans Hall, University of California, Berkeley, California 94720,
USA \printead{e1}.}

\end{aug}

%
\begin{abstract}
Residuals in regression models are often spatially correlated. Prominent
examples include studies in environmental epidemiology to understand
the chronic health effects of pollutants. I consider the effects of
residual spatial structure on the bias and precision of regression
coefficients, developing a simple framework in which to understand
the key issues and derive informative analytic results. When unmeasured
confounding introduces spatial structure into the residuals, regression
models with spatial random effects and closely-related models such
as kriging and penalized splines are biased, even when the residual
variance components are known. Analytic and simulation results show
how the bias depends on the spatial scales of the covariate and the
residual: one can reduce bias by fitting a spatial model only when
there is variation in the covariate at a scale smaller than the scale
of the unmeasured confounding. I~also discuss how the scales of the
residual and the covariate affect efficiency and uncertainty estimation
when the residuals are independent of the covariate. In an application
on the association between black carbon particulate matter air pollution
and birth weight, controlling for large-scale spatial variation appears
to reduce bias from unmeasured confounders, while increasing uncertainty
in the estimated pollution effect.
\end{abstract}

%

\end{frontmatter}

\section{Introduction}\label{sec:Introduction}

Spatial confounding is likely present in many of the applied contexts
in which residuals are spatially correlated, particularly in public
health and social science. Consider the motivating example of the\break
health effects of exposure to (spatially varying) air pollution, an
important public health issue. Many variables that explain variability
in the response, including potential confounding variables that may
be correlated with exposure, also vary spatially. For example, large-scale
regional patterns in air pollution may be correlated with regional
patterns in diet, income and other risk factors for a health outcome
of interest. Small-scale patterns in air pollution from local sources
may be correlated with risk factors as well, for example, if lower-income
people live nearer to busy roads or industrial sources. If confounding
variables are not measured, it will be difficult to distinguish the
effect of air pollution from residual spatial variation in the health
outcome. I use the term spatial confounding to characterize this situation.
Researchers have modeled the spatial structure in the outcome with
the apparent goal of reducing confounding bias
(e.g., Clayton, Bernardinelli and Montomoli, \citeyear{Clayetal1993};
\niekcite{Popeetal2002};
\niekcite{Cakmetal2003};
\niekcite{Biggetal2005}).
However, the statistical mechanism for reducing bias does not appear
to be well understood nor investigated rigorously in the statistical
or applied literature.

%
%
\begin{figure*}[b]

\includegraphics{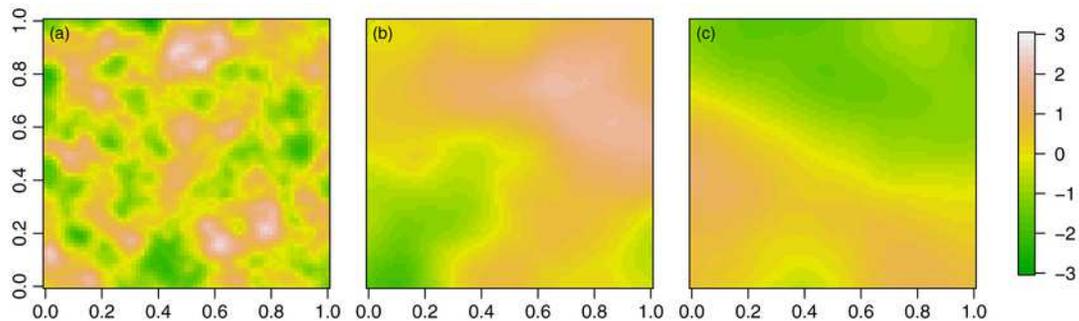}

\caption{Gaussian process realizations using the \matern covariance
(see Section
\protect\ref{sub:Analytic-Framework}) for three values of $\theta$, with
\textup{(a)} high-frequency, small (fine)-scale variability when $\theta=0.1$,
\textup{(b)} moderate scale variability when $\theta=0.5$, and
\textup{(c)} low-frequency, large-scale variability when $\theta=0.9$.}\label{fig:scales}
\end{figure*}

To consider the problem formally, start with simple linear regression
with spatial structure:
\begin{eqnarray}\label{eq:basic}
Y_{i} & = & \beta_{0}+\beta_{x}X_{i}+e_{i}, \quad   i=1,\ldots,n,\nonumber
\\[-8pt]\\[-8pt]
\bld{e} & \sim& \N(\bld{0},\mathit{\bolds{\Sigma}}),\nonumber
\end{eqnarray}
where each outcome, $Y_{i}$, is associated with a spatial location,
$\bld{s}_{i}\in\Re^{2}$. $X_{i}$ is the corresponding value of a
univariate regressor of interest, which may also vary spatially, in
which case we would represent $X_{i}$ as $X(\bld{s}_{i})$. $\bld
{e}=(e_{1},\ldots,e_{n})^{T}$
is the vector of errors, whose covariance matrix, $\bolds{\Sigma}$,
captures any residual spatial correlation, as well as independent
variation. The regression coefficients, $\beta=\{\beta_{0},\beta_{x}\}$,
are unknown, and estimation of $\beta_{x}$ is of primary interest.
Spatial statistics and regression texts note that the ordinary least
squares (OLS) estimator for $\beta_{x}$ in this setting is unbiased
but inefficient, and the usual OLS variance estimator is incorrect.
Assuming known $\bolds{\Sigma}$, the generalized least squares (GLS)
estimator is the most efficient estimator. However, little appears
to be known about how the spatial scales of the residual variability
and of $X$ affect inference. Spatial structure in $X$ is very common
in applications and complicates the problem because $X$ and the residual
spatial structure compete to explain the variability in the response
\citep{WallGotw2004}. Furthermore, it would not be surprising if
the spatial correlation in the residuals were caused by an unmeasured
spatially varying confounder; I next introduce another representation
of (\ref{eq:basic}) to enable exploration of confounding. Motivated
by the air pollution example, I~will refer to $X$ as the ``exposure.''

One can obtain the basic spatial regression model (\ref{eq:basic})
using a simple mixed model,
\begin{equation}\label{eq:random-model}
Y_{i}=\beta_{0}+\beta_{x}X(\bld{s}_{i})+g(\bld{s}_{i})+\varepsilon_{i},
\end{equation}
with random effects, $\bld{g}=(g(\bld{s}_{1}),\ldots,g(\bld{s}_{n}))^{T}$,
and\break white noise errors, $\varepsilon_{i}\stackrel{\smttt{i.i.d.}}{\sim
}\mathcal{N}(0,\tau^{2})$.
Suppose the random effects are spatially correlated, with $\bld{g}\sim\N
(\bld{0},\break\sigma_{g}^{2}\mathit{\bld{R}(\theta_{g})})$,
where $\mathit{\bld{R}(\theta_{g})}$ is a spatial correlation matrix
parameterized by $\theta_{g}$, a spatial range parameter, and $\sigma_{g}^{2}$
is the variance of the random effects. Marginalizing over $\bld{g}$
gives the marginal likelihood,%
\begin{eqnarray}\label{eq:basic-marg-lik}
\bld{Y}&=&(Y_{1},\ldots,Y_{n})^{T}\nonumber\\[-8pt]\\[-8pt]
&\sim&\mathcal{N}\bigl(\beta_{0}\bld{1}+\beta_{x}\bld{X},\sigma_{g}^{2}\mathit{\bld{R}(\theta_{g})}+\tau^{2}\bld
{I}\bigr),\nonumber
\end{eqnarray}
where $\bld{1}$ is an $n$-vector of ones, $\bld{I}$ is the identity
matrix, and $\bld{X}=(X_{1},\ldots,X_{n})^{T}$. Here $\bolds{\Sigma}$
in (\ref{eq:basic}) is explicitly decomposed into spatial and nonspatial
components. An alternative formulation would specify the unknown spatial
function, $g(\bld{s})$, as a penalized spline, where a penalty parameter
plays the role of $\{\theta_{g},\sigma_{g}^{2}\}$ in the marginal
likelihood in penalizing complexity of the spatial structure. The
exposure may itself be spatially correlated. For example, if $X(\bld{s})$\vspace*{1pt}
is a Gaussian process, then $\bld{X}\sim\N(\bld{0},\sigma_{x}^{2}\mathit{\bld{R}(\theta_{x})})$,
with parameters analogous to those for $\bld{g}$. To demonstrate
processes operating at different spatial scales, Figure~\ref{fig:scales}
shows simulated spatial surfaces as one varies the spatial range parameter,
$\theta$, in a Gaussian process model.

The spatial statistics literature assumes that the error, $e_{i}$
in (\ref{eq:basic}), is independent of the covariate(s)
(\niekcite{Cres1993}; \niekcite{WallGotw2004}),
with little or no discussion of the possibility that the error involves
variation from unmeasured confounders. Henceforth, I will refer to
the errors as residuals because of the common use of the term ``spatial
residual'' to refer to unexplained spatial variability. To explore
the possibility of confounding, let's consider $\bld{g}\equiv\beta
_{z}\bld{Z}$
to be induced as the effect, $\beta_{z}$, of an unmeasured variable,
$Z$, on the outcome. $\bld{Z}=(Z(\bld{s}_{1}),\ldots,Z(\bld{s}_{n}))^{T}$
may also be spatially correlated, for example, $\bld{Z}\sim\N(\bld
{0},\sigma_{z}^{2}\mathit{\bld{R}(\theta_{z})})$,
such that $\sigma_{z}^{2}=\break\sigma_{g}^{2}/\beta_{z}^{2}$, where $\theta_{z}$
is again a spatial\vspace*{1pt} range parameter. If $Z$ and $X$ are dependent,
then $Z$ is an unmeasured spatial confounder. Derivation of the marginal
likelihood should be done by integrating over the (unknown) conditional
distribution of $\bld{Z}$ given $\bld{X}$, whereas the integration
leading to (\ref{eq:basic}) ignores the dependence. Note that if
$\bld{X}$ and $\bld{Z}$ are considered fixed, then association between
$\bld{X}$ and $\bld{g}\equiv\beta_{z}\bld{Z}$ is known as concurvity
(\niekcite{Bujaetal1989}; \niekcite{Ramsetal2003}).

In the applied literature, practitioners often recognize the need
to consider residual spatial structure in the outcome, with language
of ``control'' or ``accounting'' for autocorrelation, and they fit models
(such as kriging or spatial random effects) that implicitly assume
independence of the residual and\break the exposure
(\niekcite{Burnetal2001};
\niekcite{Cakmetal2003};
\niekcite{Cho2003};
\niekcite{Burdetal2005};
\niekcite{Auguetal2007};
\niekcite{Molietal2007};
\niekcite{Cerdetal2009};\break
\niekcite{Leeetal2009}).
With the recent exception of \citet{HodgReic2010}, formal statements
of the goals and properties of fitting such spatial models are generally
absent. However, much of the interest appears to lie in using the
spatial residual structure to try to account for spatial confounding,
with the implicit assumption that such models reduce or eliminate
confounding bias\break
(e.g.,
\niekcite{Clayetal1993};\break
\niekcite{Popeetal2002};
\niekcite{Cakmetal2003};
\niekcite{Rich2003};
\niekcite{Biggetal2005}).
One approach is to explicitly consider the spatial scales involved,
hoping that accounting for variation at a relatively large spatial
scale allows for identification of the parameter of interest based
on exposure heterogeneity at a smaller spatial scale
(e.g.,
\niekcite{Burnetal2001};
\niekcite{Cakmetal2003};
\niekcite{Zegeetal2007}).
This smaller scale variation may be less prone to confounding in a
given application. However, this consideration of spatial scale is
often not explicit, and effects of scale on bias reduction, while
sometimes hinted at, have not been developed formally.

In the analogous context of time series modeling of air pollution,
\citet{Domietal2004} attempt to attribute all the temporally correlated
variability in the outcome to the residual term in order to identify
the effect of exposure based on the temporally uncorrelated (and presumably
unconfounded) heterogeneity in the exposure. \citet{Domietal2004}
provide no guidance in the scenario that the exposure cannot be decomposed
into autocorrelated and uncorrelated components. This issue also applies
to the approach of \citet{LombSper2007}, who filter out the dependence
between fixed and random effects. In the spatial setting, in which
measurements cannot be made at all locations, accurate estimation
of the uncorrelated component, if such a component even exists, is
rare: consider atmospheric phenomena such as temperature and air~pol\-lu\-tion.
A~common situation in which fine-scale heterogeneity is not resolved
involves prediction of spatially varying exposure values using averages
of nearby measurements or spatial smoothing techniques. Hence, I~seek
to address the problem when all of the measured components of variation
in exposure are spatial.

In this paper I address estimation in simple regression models with
spatial residual structure. I focus on the properties of penalized
models, using a simple mixed model fit by GLS, equivalent to universal
kriging, to analyze the effects of the spatially correlated residual structure
on fixed effect estimators. Section~\ref{sec:bias} focuses on bias
from spatial confounding. I report analytic results when the full
covariance structure is known and supporting simulations when the
covariance (or the amount of smoothing in penalized spline models)
is estimated from the data. I assess the use of sensitivity analysis
approaches based on spline models that explicitly consider the bias-variance
tradeoff involved in choosing the spatial scale at which to model
the residual variation. Section~\ref{sec:Precision} focuses on precision
of estimators when there is no association between exposure and residual
(no spatial confounding). I close with a case study of the effects
of air pollution on birthweight (Section~\ref{sec:Case-study:-Massachusetts}).

\section{Spatial Confounding and Bias}\label{sec:bias}

\subsection{Identifiability}\label{sec:Identifiability}

A key consideration in the basic model (\ref{eq:random-model}) is
identifiability of $\beta_{x}$ and $g(\bld{s})$. A closely-related
question is how the estimation procedure attributes variability between
the exposure and the spatial residual term (the random effects). In
the simple linear model, attribution of variability to the covariates
rather than the error term is favored because this allows the estimate
of the error variance to decrease, with the normalizing constant of
the likelihood favoring smaller error variance. In the spatial model,
if the spatial term, $\bld{g}$, is unconstrained, then $\beta_{x}\bld{X}$
and $\bld{g}$ are not identifiable in the likelihood: one could remove
the covariate from the model and redefine $g^{*}(\bld{s})\equiv\beta
_{x}X(\bld{s})+g(\bld{s})$
with no change in the likelihood. Identifiability comes through constraints
on $\bld{g}$, either by (1) penalizing lack of smoothness in $g(\bld{s})$,
(2) considering $\bld{g}$ to be a random effects term, or (3)~having
a prior on $\bld{g}$. These approaches give higher penalized likelihood,
marginal likelihood or posterior density, respectively, when variability
is attributed to the unpenalized fixed effects term rather than to
the spatial term. In the spatial confounding context this dynamic
causes bias in estimation of $\beta_{x}$, for example, as seen in
the simulations of \citet{Pengetal2006}. An alternative constraint
is to represent $\bld{g}$ in a reduced dimension basis, say, as a
regression spline. In this case the model is identifiable if there
is a component of variability in $\bld{X}$ that cannot be explained
by the spline structure, that is, if $\bld{X}$ is not perfectly collinear
with the columns of the chosen basis matrix.

\subsection{Analytic Framework}\label{sub:Analytic-Framework}

To consider bias from unmeasured spatially varying confounders, take
the following model as the data-generating mechanism,
\begin{equation}\label{eq:generative-model}
Y_{i}\sim\mathcal{N}\bigl(\beta_{0}+\beta_{x}X(\bld{s}_{i})+\beta_{z}Z(\bld
{s}_{i}),\tau^{2}\bigr),
\end{equation}
with the notation as in Section~\ref{sec:Introduction}. For each
location, $\bld{s}$, suppose the correlation of $X(\bld{s})$ and
$Z(\bld{s})$ over repeated sampling at the location is $\rho\ne0$,
so that $Z$ is a confounder. Suppose further that $Z$ is not observed
and that one models the residual spatial structure in the outcome
through spatially correlated random effects, $\bld{g}\sim\mathcal{N}(\bld{0},\sigma_{g}^{2}\mathit{\bld{R}(\theta_{g})})$
as in (\ref{eq:random-model}). Finally, suppose that one ignores
the correlation between $\bld{g}\equiv\beta_{z}\bld{Z}$ and $\bld{X}$
and integrates over the\break marginal distribution for $\bld{g}$, giving
(\ref{eq:basic-marg-lik}). Equivalently, $Y_{i}=\beta_{0}+\beta
_{x}X(\bld{s}_{i})+\varepsilon_{i}^{*}$.
The induced correlation between $X$ and $\varepsilon^{*}$ violates
the usual regression assumption that the error is independent of the
covariate, leading to bias. From the random effects perspective, we
have (incorrectly) assumed that the random effects are independent
of the covariate, a key (but often unstated) assumption of mixed effects
models (\niekcite{BresClay1993}; \ \ \niekcite{Diggetal2002}, page 170).

The treatment of $X(\bld{s})$ and $Z(\bld{s})$ as random naturally
induces spatial structure. However, in a given data set the most plausible
repeated sampling framework may suggest that $\bld{X}$ and $\bld{Z}$
reflect spatial structure that does not arise from a stochastic data
generating process. Rather, one might consider $X(\bld{s})$ and $Z(\bld{s})$
to be fixed unknown functions, particularly when $\bld{X}$ and $\bld{Z}$
vary at large scales, which mimics the partial spline/partial linear
setting. This also is consistent with the treatment of large-scale
variation in the mean term in traditional kriging. Consider the case
when there is concurvity between the two fixed functions, reflected
in a nonzero empirical correlation, $\hat{\rho},$ between $\bld{X}$
and $\bld{g}\equiv\beta_{z}\bld{Z}$ as calculated over the collection
of locations (e.g., the concurvity in\break the simulations of \
\niekcite{Ramsetal2003};
\niekcite{Heetal2006};\break
\niekcite{Pengetal2006}).
In the partial linear/partial spline setting it is well known that
such association between the exposure and the nonparametric smooth
term causes bias (\niekcite{Rice1986}, Equation~28; \niekcite{Spec1988}).
In any real data set, the orthogonality needed for $\hat{\rho}\approx0$
seems particularly unlikely if both $\bld{X}$ and $\bld{Z}$ vary
at large scale relative to the size of the domain (though $\hat{\rho}<0$
may be as much a possibility as $\hat{\rho}>0$).

The stochastic generative model is still useful under this framework
of fixed functions because realizations of $X(\bld{s})$ and $Z(\bld{s})$
give plausible values for $\bld{X}$ and $\bld{Z}$ that could arise
in real applications for which there is no reasonable stochastic mechanism.
I choose to treat $\bld{X}$ and $\bld{Z}$ stochastically, and I
use $\rho$ to quantify explicitly the strength of association between
the residual spatial variation and the exposure. This approach allows
for some simple, useful analytic results and is further justified
in that the variation that an unmeasured $\bld{Z}$ induces in $\bld{Y}$
is necessarily treated stochastically as part of the residual in actual
applications. In some cases I report results conditional on $\bld{X}$,
and in others I also average over the stochastic variability in $\bld{X}$
and over variability in the spatial locations of the observations.

Since $Z$ represents an unmeasured confounder, I assess the inferential
properties of fitting a regression model by maximizing the marginal
likelihood (\ref{eq:basic-marg-lik}) using GLS, thereby ignoring
correlation between the residual and the exposure. I assess bias as
a function of the spatial scales of $X(\bld{s})$ and $Z(\bld{s})$,
which I suppose to be generated as Gaussian processes with \matern
spatial correlation function
\[
R(d;\theta,\nu)=\frac{1}{\Gamma(\nu)2^{\nu-1}}
\biggl(\frac{2\sqrt{\nu }d}{\theta} \biggr)^{\nu}\mathcal{K}_{\nu}
\biggl(\frac{2\sqrt{\nu}d}{\theta} \biggr),
\]
where $d$ is the Euclidean distance between two locations, $\theta$
is the spatial range parameter, and $\mathcal{K}_{\nu}(\cdot)$ is
the modified Bessel function of the second kind, whose order is the
smoothness parameter, $\nu$. I fix $\nu=2$, which gives continuous
and differentiable\break Gaussian process realizations. This reflects an
assumption of some smoothness in the spatial processes under consideration,
but I also consider results based on an exponential correlation function
(i.e., $\nu=0.5$). The model~(\ref{eq:basic-marg-lik}) is equivalent
to both a mixed model and a universal kriging model if one knows the
variance and spatial dependence parameters. Furthermore, given the
extensive use of penalized splines in applications, and the connection
between penalized splines and mixed models\break \citep{Ruppetal2003},
I also consider the use of a penalized spline to represent $g(\bld{s})$.

In the nonspatial context, one would generally try to adjust for
confounding by including relevant covariates as fixed effects; in
the spatial context one could include spatial regression spline terms.
The basic question that I explore in the remainder of Section~\ref{sec:bias}
is the extent to which inclusion of a spatial random effect term or
a penalized spline can adjust for unmeasured spatial confounding,
given that these approaches do not involve a projection in the way
that a regression spline does. The random effects and penalized spline
approaches do estimate the residual spatial variation based on a bias-variance
tradeoff (e.g., Claeskens, Krivobokova and Opsomer, \citeyear{Claeetal2009}), and the penalized spline
is a regression spline in the limit as the penalty goes to zero. So
it seems plausible that these approaches may reduce bias by at least
partially adjusting for the unmeasured spatial confounder. I will
show that the spatial scales involved are critical.

\subsection{Bias with Known Parameters}\label{sub:Bias-with-known}

This section considers bias when I suppose that the variance parameters
are known and only the regression coefficients, $\beta_{0}$ and $\beta_{x}$,
are unknown. The initial results concern the situation when the exposure,
$X$, and the unmeasured confounder, $Z$, vary at the same spatial
scale. I then assess what happens when $X$ varies at two scales and
one is the same scale as the single-scale confounder. Finally, I consider
the possibility that there is additional variability in the outcome
at another scale, but uncorrelated with $X$.

To start, suppose that $X(\bld{s})$ and $Z(\bld{s})$ share the same
spatial correlation range, $\theta$, but may have different marginal
variances, namely, $\bld{X}\sim\mathcal{N}(\mu_{x}\bld{1},\break\sigma
_{x}^{2}\mathit{\bld{R}(\theta)})$
and $\bld{Z}\sim\mathcal{N}(\mu_{z}\bld{1},\sigma_{z}^{2}\mathit{\bld
{R}(\theta)})$
and $\operatorname{Cov}(\bld{X},\bld{Z})=\rho\sigma_{x}\sigma_{z}\mathit{\bld
{R}(\theta)}$.
Straightforward conditional normal calculations give%
\begin{eqnarray}\label{eq:same-scale}
\mathrm{E}(\hat{\beta}_{x}|\bld{X})
& = & \beta_{x}+ [(\bolds{\mathcal{X}}^{T}\bolds{\Sigma}^{-1}
\bolds{\mathcal{X}})^{-1}\bolds{\mathcal{X}}^{T}\bolds{\Sigma}^{-1}\mathrm{E}(\bld{Z}|\bld{X})\beta_{z}]_{2}\nonumber\\
& = & \beta_{x}+ \left[\rule{0pt}{23pt}(\bolds{\mathcal{X}}^{T}\mathit{\bolds{\Sigma}}^{-1}\bolds{\mathcal{X}})^{-1}\bolds{\mathcal{X}}^{T}\bolds{\Sigma}^{-1}
\bolds{\mathcal{X}}\nonumber
\right.
\\[-8pt]\\[-8pt]
&&{}\left.
\hspace*{54pt}\cdot
\pmatrix{
\displaystyle \mu_{z}-\rho\frac{\sigma_{z}}{\sigma_{x}}\beta_{z}\mu_{x}\vspace*{4pt}\cr
\displaystyle \rho\frac{\sigma_{z}}{\sigma_{x}}\beta_{z}
}
\right]_{2}\nonumber\\
& = & \beta_{x}+\rho\frac{\sigma_{z}}{\sigma_{x}}\beta_{z},\nonumber
\end{eqnarray}
where $\bolds{\mathcal{X}}=[\bld{1}\quad \bld{X}]$, $[\cdot]_{2}$
indicates the second element of the 2-vector, and $\bolds{\Sigma}=\sigma
_{g}^{2}\mathit{\bld{R}(\theta)}+\tau^{2}\bld{I}$.
The resulting bias, $\rho{\frac{\sigma_{z}}{\sigma_{x}}}\beta_{z}$,
is the same as if $\bld{X}$ and $\bld{Z}$ were not spatially structured
and is also equal to the bias under OLS. This demonstrates that we
have not adjusted for confounding at all by fitting the model that
includes spatial structure. As with OLS, the model attributes as much
of the variability as possible to the exposure, rather than to the
spatially correlated residual term, including all of the variability
in $\bld{Z}$ that is related to $\bld{X}$. If $\rho=0$, the bias
is zero in (\ref{eq:same-scale}). This occurs because we average
over stochastic variability in $\bld{Z}$, so any nonorthogonality
between $\bld{X}$ and $\bld{Z}$ in individual realizations contributes
to variance rather than bias. This contrasts with the bias terms in
\citet{Rice1986} and \citet{Domietal2004}, which are caused by
nonorthogonality of the fine-scale variation in $\bld{X}$ and the
nonparametric component of the model, since neither is treated stochastically.

%
%
\begin{figure*}[b]

\includegraphics{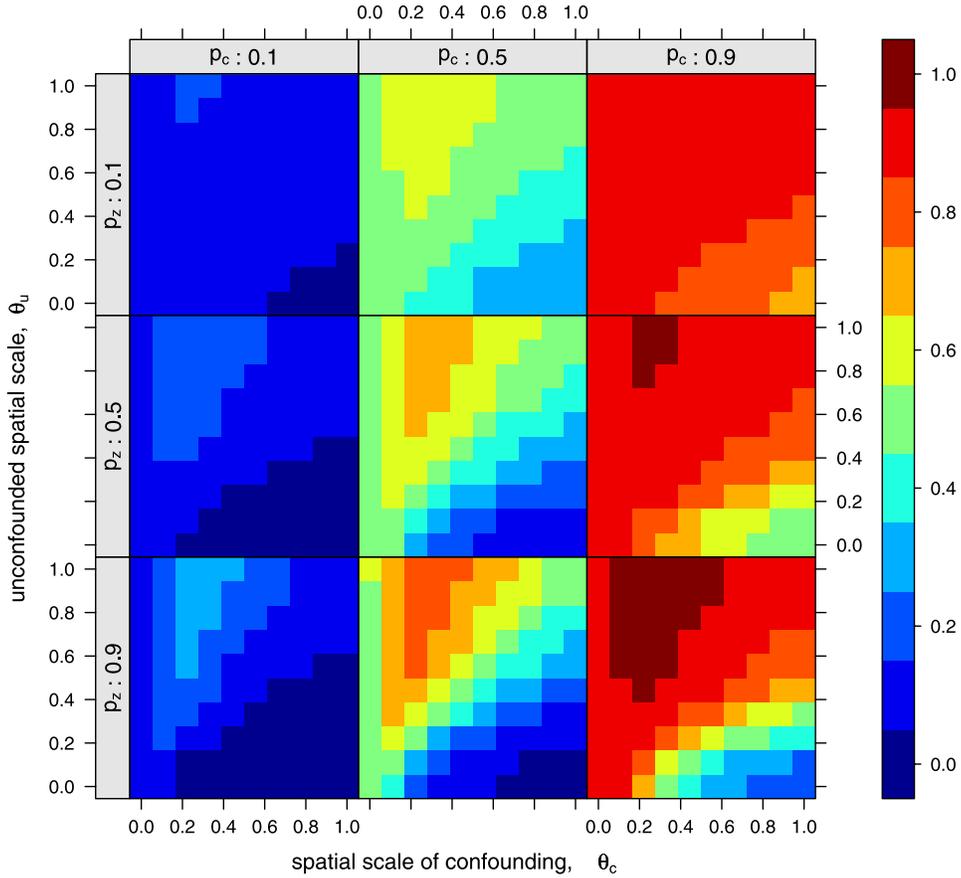}

\caption{The expected value of the bias modification term, $\hat{\mathrm{E}}_{X}k(\bld{X})$,
as a function of the spatial scales of confounded ($\theta_{c}$)
and unconfounded ($\theta_{u}$) variability for a selection of values
of $p_{z}$ and $p_{c}$. $k(\bld{X})$ quantifies the amount of bias
relative to the bias in the same-scale setting or with nonspatial
confounding ($\rho\sigma_{z}\beta_{z}/\sigma_{x}$). Along the diagonal
($\theta_{c}=\theta_{u}$) $\mathrm{E}_{X}k(\bld{X})=p_{c}$, which
is equivalent to no bias reduction. Values near zero indicate substantial
bias reduction. \label{fig:bias-exact}}
\end{figure*}

Next, I keep the same data-generating and model-fitting framework,
but explore the situation in which the exposure varies at two scales.
I suppose that $X(\bld{s})$ is a multi-scale process and introduce
correlation between $Z(\bld{s})$ and one of the components of $X(\bld{s})$.
Let $\bld{X}=\bld{X}_{c}+\bld{X}_{u}$ be decomposed into a component,
$\bld{X}_{c}$, that is at the same scale as the confounder, $\bld{Z}$,
and a component at a different scale, $\bld{X}_{u}$, which is independent
of $\bld{X}_{c}$ and $\bld{Z}$. Specifically, take $\operatorname{Cov}(\bld
{X})=\sigma_{c}^{2}\mathit{\bld{R}(\theta_{c})}+\sigma_{u}^{2}\mathit{\bld{R}(\theta_{u})}$,
$\operatorname{Cov}(\bld{Z})=\break\sigma_{z}^{2}\mathit{\bld{R}(\theta_{c})}$
and $\operatorname{Cov}(\bld{X},\bld{Z})=\operatorname{Cov}(\bld{X}_{c},\bld{Z})=\rho
\sigma_{c}\sigma_{z}\mathit{\bld{R}(\theta_{c})}$.
After some straightforward algebra and matrix manipulations, we have
\begin{eqnarray}\label{eq:diff-scale}
\mathrm{E}(\hat{\beta}_{x}|\bld{X})
& = & \beta_{x}+ [(\bolds{\mathcal{X}}^{T}\bolds{\Sigma}^{*-1}\bolds{\mathcal{X}})^{-1}\bolds{\mathcal{X}}^{T}\nonumber
\\
&&\hspace*{41pt}{}\cdot{}\bolds{\Sigma}^{*-1}\mathrm{E}(\bld{Z}|\bld{X})\beta_{z}]_{2}\\
& = & \beta_{x}+c(\bld{X})\rho\frac{\sigma_{z}}{\sigma_{c}}\beta_{z},\nonumber
\end{eqnarray}
where
\begin{eqnarray*}
k(\bld{X})&\equiv& [(\bolds{\mathcal{X}}^{T}\bolds{\Sigma}^{*-1}\bolds{\mathcal{X}})^{-1}\bolds{\mathcal{X}}^{T}\bolds{\Sigma}^{*-1}\bld{M}(\bld{X}-\mu_{x}\bld{1}) ]_{2}p_{c},
\\
\bolds{\Sigma}^{*}&\equiv&\frac{\beta_{z}^{2}\sigma_{z}^{2}\mathit{\bld
{R}(\theta_{c})}+\tau^{2}\bld{I}}{\beta_{z}^{2}\sigma_{z}^{2}+\tau
^{2}}= \bigl((1-p_{z})\bld{I}+p_{z}\mathit{\bld{R}(\theta_{c})} \bigr),
\\
\bld{M}&\equiv&\bigl(p_{c}\bld{I}+(1-p_{c})\mathit{\bld{R}(\theta_{u})}\mathit{\bld{R}(\theta_{c})}^{-1}\bigr)^{-1}
\end{eqnarray*}
and $p_{z}\equiv\beta_{z}^{2}\sigma_{z}^{2}/(\beta_{z}^{2}\sigma
_{z}^{2}+\tau^{2})$.
We see that the bias term is proportional to that in the single-scale
setting, multiplied by an additional term, $k(\bld{X})$, that modulates
the bias. $k(\bld{X})$ necessarily includes an extra multiplicative
factor, $p_{c}\equiv\sigma_{c}^{2}/(\sigma_{c}^{2}+\sigma_{u}^{2})$,
that quantifies the magnitude of the confounded component of $\bld{X}$
relative to the total variation in $\bld{X}$. While the term $k(\bld{X})$
is complicated, we can explore its dependence on the spatial scales
($\theta_{c}$ and $\theta_{u}$) and the magnitudes of the variance
component ratios ($p_{z}$ and $p_{c}$) to see how the bias compares
to the same-scale setting. In the following results I average over
the variability in $\bld{X}$.

For a grid of $n=100$ locations on the unit square, Figure~\ref{fig:bias-exact}
shows the average of $k(\bld{X})$ over 1000 simulations as a function
of $\theta_{c}$ and $\theta_{u}$, for combinations of $p_{c}$ and
$p_{z}$, where the empirical average approximates the expectation
with respect to the distribution of $\bld{X}$. There is a simple
pattern to the bias modification relative to the same-scale setting.
For $\theta_{c}=\theta_{u}$ (the diagonal elements on the $1\dvtx 1$ line),
we do not need simulation: $\mathrm{E}_{X}k(\bld{X})=p_{c}$, which
is equivalent to the same-scale result (\ref{eq:same-scale}), after
accounting for the proportion of variability in $\bld{X}$ that is
confounded, $p_{c}$. Note that if one estimates the analogous bias
to (\ref{eq:diff-scale}) for OLS applied to spatial data, it is nearly
constant regardless of the spatial scales [$\hat{\mathrm{E}}_{X}k(\bld
{X})\approx p_{c}$;
not shown]. Only when $\theta_{u}<\theta_{c}$, and particularly when
$\theta_{u}\ll\theta_{c}$, do we see less bias than in the same-scale
setting, with clear potential for bias reduction from modeling the
residual spatial variation in the outcome (recall Figure~\ref{fig:scales}
to interpret the values of $\theta$). Above\vspace*{1pt} the diagonal, for $\theta
_{u}>\theta_{c}$,
$\hat{\mathrm{E}}_{X}k(\bld{X})>p_{c}$, indicating more bias when the
scale of confounding is smaller than the scale of unconfounded variability.
This situation may be of limited practical interest, because it's
not clear that there are real applications in which the unconfounded
variation in the exposure occurs at larger scales than the confounded
variation. However, it does show that there are circumstances in which
bias is larger than under OLS, a point also made in \citet{HodgReic2010}.
Note that the patterns in Figure~\ref{fig:bias-exact} are qualitatively
similar regardless of the values of $p_{c}$ and $p_{z}$. Quantitatively,
for larger values of $p_{c}$, corresponding to a larger proportion
of the variation in the exposure being confounded variation, bias
is larger. For larger values of $p_{z}$, corresponding to a larger
proportion of the residual variation being the contribution of the
confounder, the effects of the spatial scales are more distinct. The
results highlight that inclusion of the spatial residual does not
give unbiased estimates and bias is substantial in many scenarios
even when the covariance parameters are known. Results are very similar
when I sample locations uniformly on the unit square or in a clustered
fashion (using a Poisson cluster process).

%
%
\begin{figure*}

\includegraphics{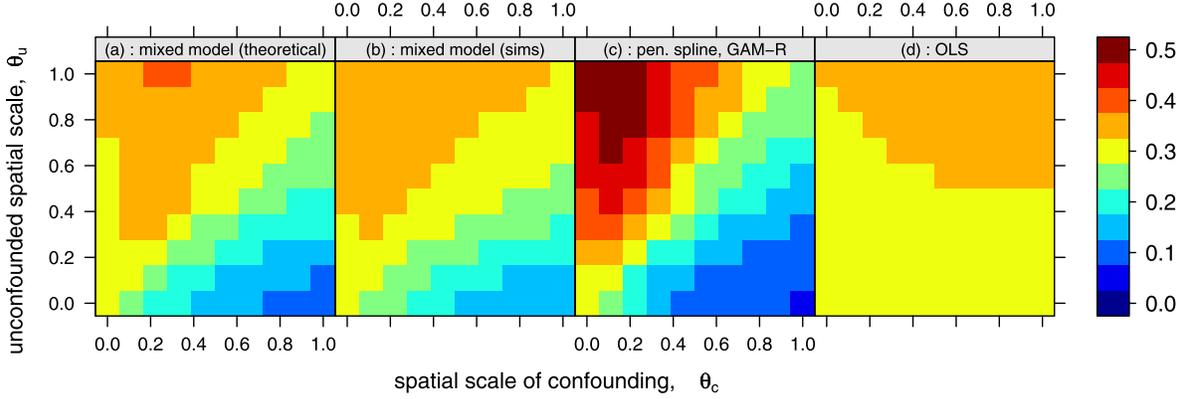}

\caption{Relative bias, $(\hat{\mathrm{E}}(\hat{\beta}_{x})-\beta_{x})/\beta_{x}$,
as a function of the spatial scales of confounded ($\theta_{c}$)
and unconfounded variability ($\theta_{u}$): \textup{(a)}~theoretical bias
for the mixed/kriging model with known variance parameters, and \textup{(b),
(c), (d)} simulated bias with estimated variance/penalty parameters for
\textup{(b)} the mixed model, \textup{(c)} a penalized spline model, and \textup{(d)} OLS.}\label{fig:sims-bias}
\end{figure*}

%
%
\begin{figure*}[b]

\includegraphics{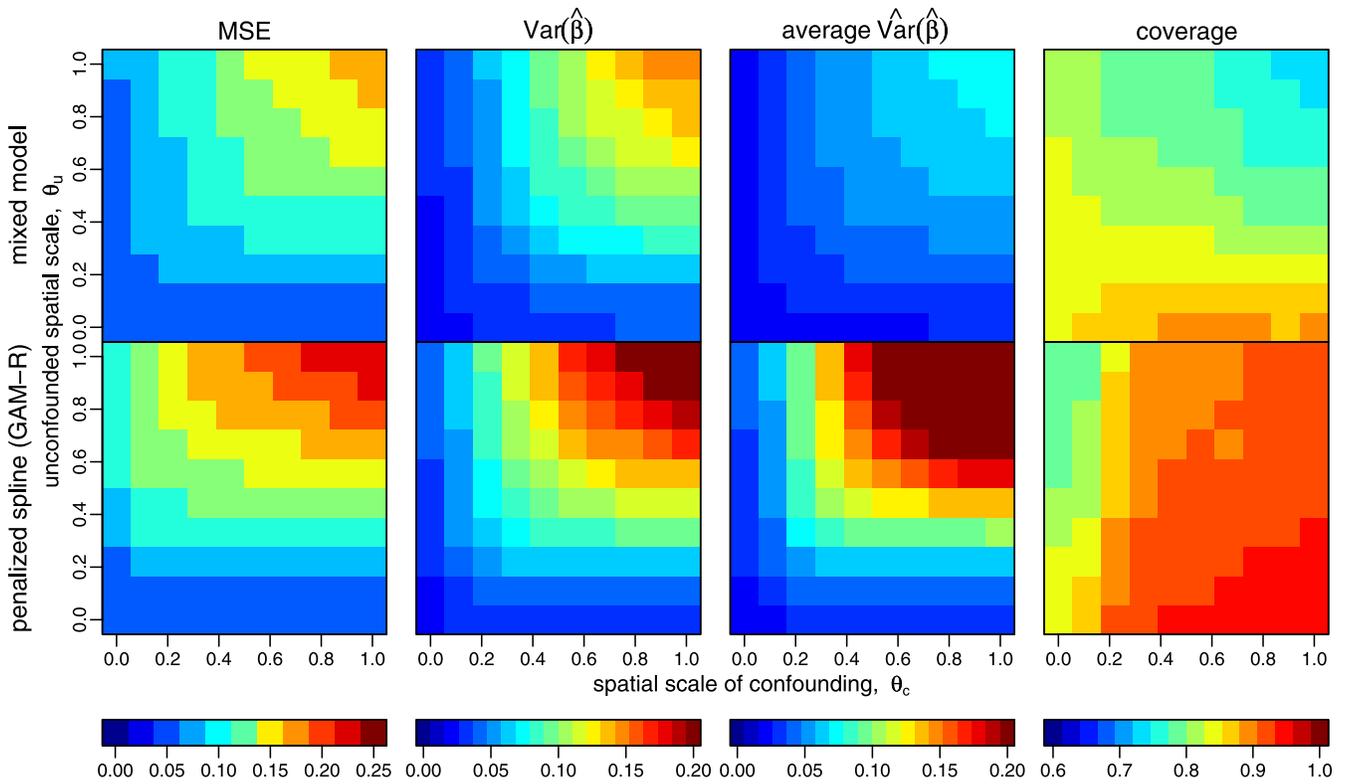}

\caption{Simulation results for (top row) mixed model/kriging fit and (bottom
row) penalized spline model. Each plot shows results as a function
of the spatial scales of the confounded ($\theta_{c}$) and unconfounded
variability ($\theta_{u}$), with MSE (first column), variance of
the estimates over the simulations (second column), average squared
standard error (third column) and coverage (fourth column).}\label{fig:sims}
\end{figure*}

The results in Figure~\ref{fig:bias-exact} correct for the complication
that the sample variance of spatial process values (calculated over
the domain) decreases as $\theta$ increases. This occurs because
the sample variance over the domain in a single spatial replicate
underestimates population variability; see Figure~\ref{fig:scales}(b)--(c)
for examples. I want to have fixed ratios of average sample variances,\vspace*{1pt}
$p_{z}\equiv\beta_{z}^{2}\mathrm{E}_{Z}s_{z}^{2}/(\beta_{z}^{2}{\mathrm{E}_{Z}s}_{z}^{2}+\tau^{2})\in\{0.1,0.5,0.9\}$
and $p_{c}\equiv\mathrm{E}_{X_{c}}s_{c}^{2}/(\mathrm{E}_{X_{c}}s_{c}^{2}+\mathrm{E}_{X_{u}}s_{u}^{2})\in\break\{0.1,0.5,0.9\}$,
for all values of $\theta_{c}$ and $\theta_{u}$, thereby avoiding
the introduction of artifacts caused solely by having ratios of sample
variances change with the spatial\vspace*{1pt} ranges. Here $s_{z}^{2}$, $s_{c}^{2}$
and $s_{u}^{2}$ are the sample variances of $\bld{Z},$ $\bld{X}_{c}$
and $\bld{X}_{u}$, respectively.\vspace*{1pt} To achieve this, I generate $\bld
{X}_{c}\sim\N(\bld{0},d_{c}^{2}\sigma_{c}^{2}\mathit{\bld{R}(\theta_{c})})$
and $\bld{X}_{u}\sim\N(\bld{0},d_{u}^{2}\sigma_{u}^{2}\mathit{\bld
{R}(\theta_{u})})$
and modify the calculation of $k(\bld{X})$ in (\ref{eq:diff-scale})
accordingly. $d_{c}$ and $d_{u}$ are functions of $\theta_{c}$
and $\theta_{u}$, respectively, that are chosen such that $\mathrm{E}_{X_{c}}s_{c}^{2}(\theta_{c})\approx\sigma_{c}^{2}$
and $\mathrm{E}_{X_{u}}s_{u}^{2}(\theta_{u})\approx\sigma_{u}^{2}$,
where $s_{c}^{2}(\theta_{c})$ is the sample variance of $\bld{X}_{c}$
for a given realization under $\theta_{c}$ and analogously for
$s_{u}^{2}(\theta_{u})$.
The expectations are taken with respect to the distribution of the
subscripted random vector. These manipulations allow me to present
bias for scenarios that correspond to specific ratios of average sample
variability of $\bld{X}_{c}$, $\bld{X}_{u}$, $\bld{Z}$ and $\bolds{\varepsilon}$
over the spatial domain.

To have only a single scale of residual spatial variability is not
very realistic. Therefore, I carried out an additional simulation
study with residual spatial variability in the outcome that is independent
of the exposure and at a smaller scale than the\break scale of $Z(\bld{s})$.
I~suppose that the data-generating model is%
\begin{equation}
\bld{Y}=\beta_{0}\bld{1}+\beta_{x}\bld{X}+\beta_{z}\bld{Z}+\bld{h}+\bolds{\varepsilon},\label{eq:multiscale-resid}
\end{equation}
and that $\bld{h}\sim\N(\bld{0},\sigma_{h}^{2}\mathit{\bld{R}(\theta_{u})})$,
independent of $\bld{X}$, $\bld{Z}$ and $\bolds{\varepsilon}$, with
all of the other details as before. Under this data-generating model
and again supposing that all variance parameters are known, simulation
estimates of $\mathrm{E}_{X}k(\bld{X})$ indicate that bias is somewhat
smaller than that seen in Figure~\ref{fig:bias-exact} for $\theta
_{c}>\theta_{u}$
and somewhat larger for $\theta_{c}<\theta_{u}$ (not shown). Note
that if the additional small-scale variability is correlated with
the exposure, then one is back in the situation of having common scales
for the exposure and the confounder, which is considered at the beginning
of this section.

\hspace*{3pt}
\subsection{Bias and Precision with Estimated Parameters}\label{sub:sims}

To generalize the results of Section~\ref{sub:Bias-with-known}, which
supposed known variance and spatial dependence parameters, I set up
a simulation study to assess the impact of estimating those parameters.
In addition to maximum likelihood estimation of a mixed effects/kriging
model based on the marginal likelihood (\ref{eq:basic-marg-lik}),
I consider the use of penalized likelihood to fit the model~(\ref{eq:random-model})
with a penalized thin plate spline spatial term for $g(\bld{s})$.
I implemented the penalized spline using gam() in R, which uses generalized
cross-validation (GCV) for data-driven smoothing parameter estimation
\citep{Wood2006}. For the core simulations, I set the following
parameter values, $\sigma_{u}^{2}=\sigma_{c}^{2}=\beta_{z}^{2}\sigma_{z}^{2}=1$,
$\tau^{2}=4$, $\beta_{x}=0.5$, $\rho=0.3$, and sample 100 spatial
locations uniformly from the unit square. For a range of values of
$\theta_{c}$ and $\theta_{u}$, I simulate 2000 data sets for each
pair \{$\theta_{c},\theta_{u}$\}. For each simulated data set, I generate
new spatial locations and new values of $\bld{X}$ and $\bld{Z}$;
I then generate $\bld{Y}$ using (\ref{eq:generative-model}). Again
we have to account for the reduced empirical spatial variability as
$\theta$ increases; these simulations have effective values of $p_{c}=0.5$
and $p_{z}=0.2$.

With regard to bias, the simulation results for the mixed/kriging
model [Figure~\ref{fig:sims-bias}(b)] reasonably match the theoretical
values with known variance parameters [Figure~\ref{fig:sims-bias}(a)].
However, when $\theta_{u}\ll\theta_{c}$, the bias is generally larger
than with known variance parameters, because the fitted model sometimes
estimates little or no spatial structure in the residuals, pushing
bias results toward the larger bias seen under OLS [Figure~\ref{fig:sims-bias}(d)].
Results for the penalized spline model [Figure~\ref{fig:sims-bias}(c)]
show smaller bias for $\theta_{u}<\theta_{c}$ than the mixed model,
presumably caused by the difference between estimating the amount
of smoothing by GCV compared to maximum likelihood. In either case,
spatial scales are critical, and bias is smaller than with OLS only
when the scale of confounding is larger than the scale of the unconfounded
variability. Additional simulations indicate that as the correlation
of confounder and exposure increases, or the magnitude of variation
in the confounder increases, or the effect size decreases, relative
bias increases (not shown). In such scenarios, substantial bias reduction
occurs only for very small spatial scales in the exposure and large
scales of confounding.

Figure~\ref{fig:sims} compares the mixed model with the penalized
spline in the context of a bias-variance tradeoff. There is a substantial
bias-variance tradeoff, with the smaller bias of the penalized spline
model (for $\theta_{c}<\theta_{u}$) trading off for increased variance.
The result is increased mean squared error (MSE) in $\hat{\beta}_{x}$,
except when $\theta_{u}$ is very small. Both model variance estimates
(third column) understate the variability in the coefficient estimates
(second column), with particular underestimation of uncertainty and
low coverage for the mixed/kriging model, and with lower coverage
as one moves away from the region of $\theta_{c}\gg\theta_{u}$. Of
course the bias causes much of the poor coverage.

Fitting the mixed/kriging model by restricted maximum likelihood (REML)
rather than maximum likelihood produces moderate improvement in coverage,
with the average variance estimate more similar to the variance of
the estimated coefficients. Using $\nu=0.5$ (i.e., an exponential
spatial correlation function) in the fitting rather than the true
$\nu=2$ has little effect on results. However, when I generate the
unconfounded variability, $\bld{X}_{u}$, based on $\nu=0.5$, bias
is substantially smaller than the core results (particularly note
that there is reduced bias relative to OLS when $\theta_{c}=\theta_{u}$),
apparently because the nondifferentiable sample paths of processes
with exponential covariance play the role of very fine-scale, unconfounded
variability. There is little change in results when using spatial
locations simulated using a Poisson cluster process with an average
of seven children per cluster and cluster kernel standard deviation
of $0.03$. Finally, simulations with $\rho=0$, that is, no confounding,
indicate no bias for either model, as expected.

Our bias results when $\rho\ne0$ are analogous to the bias seen with
penalized spline models in \citet{Heetal2006} and \citet{Pengetal2006}.
There, concurvity (i.e., $\hat{\rho}\ne0$) between the smooth temporal
term (analogous to our spatial residual) and the exposure emerged
from the fixed basis coefficients chosen based on empirical data\break examples
(R. Peng, personal communication;\break \niekcite{Heetal2006}). Similar
results are seen in the spatial settings of \citet{Ramsetal2003}.

The presence of small-scale independent variation in the residual
(\ref{eq:multiscale-resid}) reduces bias for $\theta_{u}<\theta_{c}$
(not shown), relative to the results presented above. This occurs
through an increase in the number of degrees of freedom estimated
from the data to capture residual variability, that is, undersmoothing
with respect to the variation at the $\theta_{c}$ scale, analogous
to undersmoothing in the partial spline setting
(\niekcite{Rice1986}; \niekcite{Spec1988}).
This scenario seems quite likely in applications: if there is large-scale
residual spatial structure, there is likely to be finer-scale structure
as well. Thus, analyses that attempt to best fit the data may in the
process reduce bias from confounding at the larger scales.

\subsection{The Bias-Variance Tradeoff}\label{sub:Bias-variance}

We have seen that even when all covariance parameters are known and
the scale of confounding is much larger than the scale of unconfounded
variability in $\bld{X}$, bias remains, albeit at a much reduced
level. In principle, if the structure at the confounded scale could
be exactly fit using a set of basis\break functions, such as a regression
spline (e.g.,\break \niekcite{Domietal2004}), then the exposure effect
estimate would be unbiased, as in any multiple regression. The partial
residual kernel smoothing approach of \citet{Spec1988} reduces bias
in a similar fashion, albeit without using a projection, through the
technique of twicing. However, in a real application, one has to
choose the basis functions, and if the basis functions do not fully
explain the confounded, large-scale variability, even with a basis
of seemingly sufficient dimension, this will induce a bias. One could
instead consider a penalized spline approach with penalty parameter
chosen in advance to give the desired effective degrees of freedom
(e.d.f.). For fixed e.d.f., since the penalized spline smoother is not a
true projection (\niekcite{Spec1988}; \niekcite{Pengetal2006}), one would expect
the penalized spline approach to have more bias than the regression
spline approach. Heuristically, bias in this approach occurs because
the estimated spatial term does not fully explain the confounded component
of the variability in the outcome, causing a bias analogous to that
seen in the partial spline setting (\niekcite{Rice1986}; \niekcite{Spec1988}).
However, we would expect the penalized spline to be less sensitive
to the exact form of the basis functions and number and placement
of knots, as is seen in the example (Section~\ref{sec:Case-study:-Massachusetts}).
Furthermore, one can always undersmooth to reduce the bias, following
the recommendation in the partial spline literature (\niekcite{Rice1986}; \niekcite{Spec1988}).
Thus, using a penalized spline seems reasonable, albeit without the
clean interpretation of a projection. I show below that simulations
comparing regression spline and penalized spline models support these
theoretical results from the literature, in the spatial context considered
here, with the regression spline having reduced bias and increased
variance relative to penalized modeling.

%
%
\begin{figure*}[b]

\includegraphics{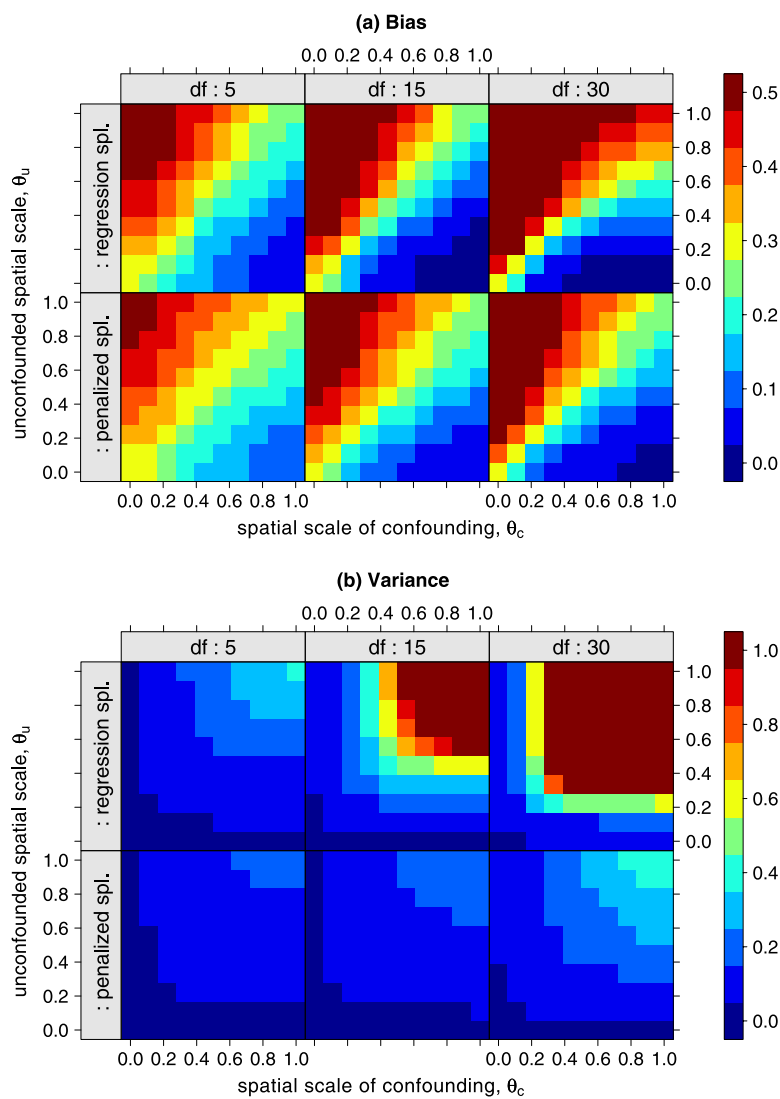}

\caption{Simulation results for relative bias \textup{(a)} and variance \textup{(b)} of
$\hat{\beta}_{x}$
as a function of the spatial scales of confounded ($\theta_{c}$)
and unconfounded ($\theta_{u}$) variability for regression splines
(top rows) and penalized splines (bottom rows) with 5, 15 and 30 e.d.f.,
where the e.d.f. are pre-specified, rather than estimated based on the
data.}\label{fig:sims-fixdf}
\end{figure*}

The primary issue in an application is choosing the amount of smoothing
to reduce bias, since inference about $\beta_{x}$ is the goal rather
than best fitting the data. Data-driven smoothing might reduce bias
(if there is small-scale residual correlation) or might have little
effect on bias (if the data suggest only large-scale residual correlation).
Thus, the reduction in bias will depend on the scales involved and
the actual amount of smoothing done, and the analysis will reveal
little about the sensitivity of estimation to scale. Instead, one
could explicitly assess the bias-variance tradeoff by varying the
amount of smoothing and assessing the sensitivity of the exposure
effect inference. One approach is a spatial analogue to the sensitivity
analysis approaches of \citet{Pengetal2006}: fit a model with spatial
basis functions and vary the e.d.f. (e.g., \niekcite{Zegeetal2007}).
Plotting $\hat{\beta}_{x}$ and uncertainty intervals as a function
of e.d.f. (or some other metric) provides an assessment of the robustness
of results to potential spatial confounding at various scales. If
one is concerned about confounding at a particular scale, then one
can report the results for an e.d.f. that would undersmooth with respect
to that scale to reduce bias, accepting the tradeoff of increased
uncertainty.

Motivated by this analysis strategy, I set up a simulation under the
settings of Section~\ref{sub:sims}, using a regression spline (i.e.,
unpenalized fixed effects) and varying the e.d.f. by changing the dimension
of the basis in gam() in R. Figure~\ref{fig:sims-fixdf}(a) shows relative
bias as a function of the spatial scales involved. As before, I focus
on the results below the $1\dvtx 1$ diagonal ($\theta_{c}=\theta_{u}$), as
this is the scenario of practical interest. By choosing a large number
of e.d.f., one can decrease bias more effectively than when estimating
the amount of smoothing from the data [i.e., Figure~\ref{fig:sims-bias}(c)].
However, with moderate and large scale variability, the variance of
the estimates in this fixed effects model increases dramatically
[Figure~\ref{fig:sims-fixdf}(b)].
This causes a concordant increase in the MSE (not shown), highlighting
the bias-variance tradeoff. In contrast, using a penalized spline
with fixed e.d.f. [fixing the smoothing parameter in gam() in R] shows
much more stable results. As expected, for a given e.d.f. bias is not
reduced as much as with a regression spline [Figure~\ref{fig:sims-fixdf}(a)],
but there is much less variability [Figure~\ref{fig:sims-fixdf}(b)].

A diagnostic approach to understanding whether the residual may include
variation from an unmeasured confounder is to assess the correlation
between the residual and the exposure. Not knowing $\beta_{x}$, one
might use a variety of plausible values of $\beta_{x}$ to estimate
$\bld{g}$ and then calculate the correlation with $\bld{X}$ (and
potentially with filtered versions of $\bld{X}$ that exclude small-scale
variation).

\subsection{Accounting for Residual Spatial Correlation}\label
{sub:Accounting-for-residual}

If one accounts for large-scale variation as a means of reducing bias,
there may still be small-scale residual variation, such as fine-scale
correlation in health outcomes related to residential sorting. As
I have shown, one can reduce potential confounding bias from this
fine-scale variation through explicit spatial modeling only if there
is variability in the exposure at an even smaller scale. If there
is not, then one is effectively assuming that the fine-scale variation
is uncorrelated with the exposure. Given this assumption, one may
need to account for the fine-scale residual spatial variation so that
uncertainty estimation for $\beta_{x}$ is not compromised (but note
the results of Section~\ref{sub:prec-ques3}). One possibility would
be to use an analysis robust to misspecification of the residual variance,
for example, using an estimating equation with uncertainty based on
the sandwich estimator, with regression spline terms in the mean to
account for large-scale spatial confounding bias. Alternatively, one
could fit a penalized model with the amount of smoothing determined
from the data. This has two effects. First, it naturally accounts
for the effect of the spatial structure on uncertainty estimation.
Second, in the presence of small-scale residual variability, the model
will naturally undersmooth with respect to large-scale variability
that may cause confounding, thereby reducing bias from confounding
at the larger scale, as discussed previously.

\section{Spatially Correlated Residuals and~Precision}\label{sec:Precision}

In this section I suppose that the residual and the exposure are independent
($\rho=0$ in the framework of Section~\ref{sec:bias}), which results
in unbiased estimation of $\beta_{x}$. I consider effects of spatial
scale on the following questions about efficiency of estimators for
$\beta_{x}$ (henceforth simply $\beta$) and quantification of uncertainty:
\begin{enumerate}[(3)]
\item[(1)] Given a fixed amount of residual variation, how is efficiency
affected by the proportion of that variation that is spatial?

\item[(2)] What is the magnitude of the improvement in efficiency when accounting
for residual spatial variation, relative to OLS?

\item[(3)] If one uses the naive estimator for the variance of the OLS estimator,
$\hat{\beta}_{\smttt{OLS}}$, what is the magnitude of the error in
uncertainty estimation compared to the correct variance estimator
for $\hat{\beta}_{\smttt{OLS}}$?
\end{enumerate}

The first question does not appear to have been raised in the literature.
With regard to the second, while we know that GLS is the most efficient
estimator when the residuals are correlated, here I investigate the
magnitude of this efficiency advantage as a function of the spatial
scales involved. Regarding the third, the conventional wisdom in the
statistical and applied literature appears to be that not accounting
for spatial structure leads to underestimation of uncertainty
(e.g., \niekcite{Lege1993}; \niekcite{Burnetal2001}; \niekcite{SchaGotw2005}, page~324). However,
I have not seen a formal quantification of this underestimation for
a regression coefficient, in contrast to our understanding of the
potentially severe underestimation of uncertainty for the mean\break of
a spatial process (\niekcite{Cres1993}, Section~1.3;\break \niekcite{SchaGotw2005}, Section~1.5).

Note that there are three variance estimators (i.e., estimators for
the sampling variability of the estimated regression coefficient)
under consideration here: the true GLS variance estimator, and the
true and naive OLS variance estimators. When $\rho=0$, OLS is unbiased,
so it makes sense to consider OLS for estimation, provided we adjust
the usual OLS variance estimator to account for the residual spatial
correlation. While actual applications will likely involve more complicated
modeling, consideration of these questions in this simple setting,
and with known variance components, helps to understand the basic
issues.

%
%
\begin{figure*}[b]

\includegraphics{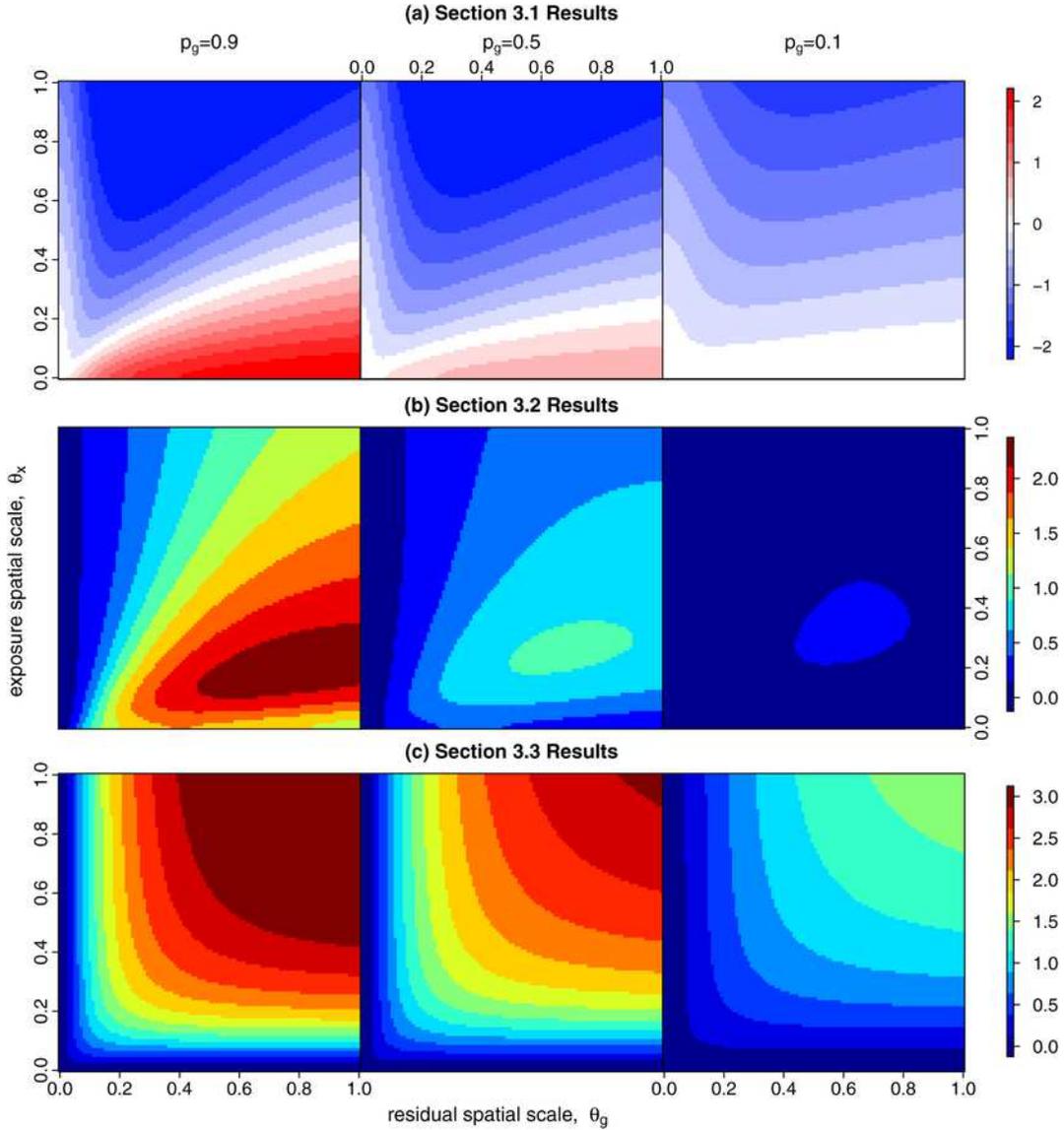}

\caption{Efficiency and precision results for three values of
$p_{g}=\sigma_{g}^{2}/(\sigma_{g}^{2}+\tau^{2})$
(columns) as a function of the spatial scales of the residual ($\theta_{g}$)
and the exposure ($\theta_{x}$). \textup{(a)} The log of the expected precision
of the GLS estimator (\protect\ref{eq:gls-prec}), relative to the expected
precision in the nonspatial setting with equivalent total residual
variation. \textup{(b)} Relative efficiency of GLS and OLS estimation, quantified
as the log of the expected ratio of GLS to OLS precision. \textup{(c)} The
log of the expected ratio of the correct and naive OLS variance estimators
(\protect\ref{eq:sec3.3}). The results are based on 500 simulations for each
set of parameter values, with a \matern correlation with $\nu=2$
and 100 locations sampled uniformly over the unit square.}\label{fig:prec-GLS}
\end{figure*}

\subsection{Relationship Between Spatial Scale\break and GLS Efficiency}\label{sub:prec-ques1}

Given a fixed amount of residual variation, how is efficiency affected
by the proportion of variation that is spatial? I quantify efficiency
in terms of precision rather than variance, as this allows for closed
form derivations.
\begin{lemma}\label{LEM}
Consider the model (\ref{eq:basic-marg-lik}) and suppose
that all parameters are known except $\beta_{0}$ and $\beta\equiv\beta_{x}$.
The expectation of the precision of $\hat{\beta}_{\smttt{GLS}}$,
with respect to the sampling distribution of $\bld{X}$, is
\begin{eqnarray}\label{eq:gls-prec}
&&\mathrm{E}_{X}(\operatorname{Var}(\hat{\beta}_{\smttt{GLS}})^{-1})\nonumber
\\
&&\quad=\frac{\sigma_{x}^{2}}{\tau^{2}+\sigma_{g}^{2}} \biggl(\operatorname{tr}\{\mathit{\tilde{\bolds{\Sigma}}}{}^{-1}\mathit{\bld{R}}(\theta_{x})\}
\\
&&\qquad\hspace*{43pt}{}-\frac{\bld{1}^{T}\mathit{\tilde{\bolds{\Sigma}}}{}^{-1}\mathit{\bld{R}}(\theta_{x})\mathit{\tilde
{\bolds{\Sigma}}}{}^{-1}\bld{1}}{\bld{1}^{T}\mathit{\tilde{\bolds{\Sigma
}}}{}^{-1}\bld{1}} \biggr),\nonumber
\end{eqnarray}
where $\mathit{\tilde{\bolds{\Sigma}}}\equiv(1-p_{g})\bld{I}+p_{g}\mathit{\bld{R}}(\theta_{g})$,
$p_{g}\equiv\sigma_{g}^{2}/(\sigma_{g}^{2}+\tau^{2})$, and the remaining
notation follows that in previous sections. See the \hyperref[append]{Appendix} for the
proof.
\end{lemma}

Note that the term in parentheses is an effective sample size, analogous
to $n-1$ in the nonspatial problem. Here the adjustment is for spatial
structure in residual and exposure, with the second component in the
parentheses analogous to the degree of freedom lost for estimating
a mean.

Figure~\ref{fig:prec-GLS}(a) shows Monte Carlo estimates of the expected
precision as a function of $\theta_{x}$ and $\theta_{g}$, averaging
(\ref{eq:gls-prec}) over 500 sets of $n=100$ locations simulated
uniformly on the unit square. I report the expected precision divided
by a baseline of $\sigma_{x}^{2}(n-1)/(\tau^{2}+\sigma_{g}^{2})$,
which is the expected precision in the nonspatial setting, supposing
that the total residual variation, $\tau^{2}+\sigma_{g}^{2}$, remains
constant. Compared\vspace*{1.5pt} to the nonspatial setting, lower precision occurs
unless the exposure varies at small spatial scale. When the exposure
varies at small spatial scale and the residual at larger spatial scale,
precision can be substantially greater than in the nonspatial setting.
The model is able to account for part of the residual variance through
the spatial structure, as if the spatial structure were an additional
covariate to which variation in the response is attributed. GLS implicitly
estimates the process, $\bld{g}$ in (\ref{eq:random-model}), that
gives rise to the marginalized model (\ref{eq:basic-marg-lik}). This
reduces the remaining ``unexplained'' residual variability and thereby
improves efficiency relative to having independent errors but equivalent
overall residual variability. In contrast, when $\bld{X}$ varies
only at large spatial scales, then efficiency decreases because of
difficulty in distinguishing $\beta X(\bld{s})$ from $g(\bld{s})$.
Results are similar using points on a regular grid or clustered based
on a Poisson cluster process.

\subsection{Efficiency of GLS and OLS Estimators}\label{sub:prec-ques2}

Here I consider how spatial scale affects the relative efficiency
of spatial and nonspatial estimators, comparing the precisions of
the OLS and\break GLS estimators. Since the true OLS variance,\break
$ [(\bolds{\mathcal{X}}^{T}\bolds{\mathcal{X}})^{-1}(\bolds{\mathcal{X}}^{T}(\tau
^{2}\bld{I}+\sigma_{g}^{2}\mathit{\bld{R}_{g}})\bolds{\mathcal{X}})(\mathit{\bolds{\mathcal{X}}}^{T}\bolds{\mathcal{X}})^{-1} ]_{2,2}$,
is a\break complicated function, it is difficult to derive closed form expressions
for efficiency relative to the GLS estimator. Instead I conduct a
small simulation study. For a regular grid of values of $\theta_{x}$
and $\theta_{g}$, I carry out 500 simulations for each pair of values,
with $n=100$ observations whose spatial locations are drawn uniformly
over the unit square domain. Note that I consider the ratio of the
GLS precision to the OLS precision,\vspace*{1pt} so the values of $\sigma_{x}^{2}$
and $\sigma_{g}^{2}+\tau^{2}$ cancel out of the ratio and do not
affect the results.

Figure~\ref{fig:prec-GLS}(b) shows the Monte Carlo estimates of the
expected relative precision, as a function of the spatial scales,
$\theta_{g}$ and $\theta_{x}$, and the proportion of the residual
variability that is spatial. When little of the residual variability
is spatial ($p_{g}=0.1$), there is little gain in precision, as expected.
When more is spatial, the gains in precision are small when $\bld{g}$
varies at a small scale, but substantial when $\bld{g}$ varies at
a large scale. Unfortunately, this is also precisely the case in which
one would be concerned about spatial confounding. If we suppose that
the large-scale structure in the residual has been controlled for
in an effort to reduce the potential for bias, then with the remaining
residual variability being fine scale, there is limited gain in precision
regardless of the spatial scale of the exposure. With locations on
a regular grid, the gains in precision are slightly less for small
values of $\theta_{g}$, while with Poisson cluster process sampling,
the gains are somewhat larger for small values of $\theta_{g}$. See
also \citet{Dowetal1982} for similar simulation results when a
Markov random field structure induces the correlation.

\subsection{Underestimation of Uncertainty by the Naive OLS Variance
Estimator}\label{sub:prec-ques3}

Applied analyses often ignore residual spatial correlation, raising
the question of how strongly uncertainty estimates are affected. One
can express the ratio of the true OLS variance to the incorrect naive
OLS variance as follows. First define $\bld{W}\equiv(\bld{X}-\bar{X}\bld{1})/s$
where $s^{2}\equiv\frac{1}{n}\sum(X_{i}-\bar{X})^{2}$. After expressing
$\hat{\beta}_{x}=[(\mathit{\bolds{\mathcal{X}}}^{T}\mathit{\bolds{\mathcal{X}}})^{-1}\bolds{\mathcal{X}}^{T}\bld{Y}]_{2}=(\tilde{\mathbf{X}}{}^{T}\bld
{\tilde{X}})^{-1}\tilde{\mathbf{X}}{}^{T}\bld{Y}$
where $\tilde{\mathbf{X}}=\bld{X}-\bar{X}\bld{1}$, we have
\begin{eqnarray*}
\frac{\operatorname{Var}_{\smttt{true}}(\hat{\beta}_{x})}{\operatorname{Var}_{\smttt
{naive}}(\hat{\beta}_{x})}&=&\frac{(\sigma_{g}^{2}+\tau^{2})^{-1}(\bld
{\tilde{X}}^{T}\tilde{\mathbf{X}})}{(\sigma_{g}^{2}+\tau^{2})^{-1}(\bld
{\tilde{X}}^{T}\tilde{\mathbf{X}})(\tilde{\mathbf{X}}{}^{T}\mathit{\tilde{\bld
{\Sigma}}}\tilde{\mathbf{X}})^{-1}\tilde{\mathbf{X}}{}^{T}\tilde{\mathbf{X}}}
\\
&=&\frac{\tilde{\mathbf{X}}{}^{T}\mathit{\tilde{\bolds{\Sigma}}}\tilde{\mathbf{X}}}{\bld
{\tilde{X}}^{T}\tilde{\mathbf{X}}}=\frac{1}{n}\bld{W}^{T}\mathit{\tilde
{\bolds{\Sigma}}}\bld{W}.
\end{eqnarray*}
Averaging over the sampling distribution of $\bld{X}$, we have
\begin{equation}\label{eq:sec3.3}
\mathrm{E}_{X}\biggl(\frac{1}{n}\bld{W}^{T}\mathit{\tilde{\bolds{\Sigma}}}\bld
{W}\biggr)=\frac{1}{n}\operatorname{tr}(\mathit{\tilde{\bolds{\Sigma}}}\operatorname{Cov}(\bld
{W})).
\end{equation}
So for $\mathit{\tilde{\bolds{\Sigma}}}\approx\bld{I}$ or $\operatorname{Cov}(\bld
{W})\approx\mathit{\bld{I}}$,
that is, when either $\theta_{g}$ or $\theta_{x}$ is close to zero,
we expect the ratio to be near one. Note also that with spatial correlation
functions that are nonnegative, the only negative contribution to
the ratio can be from negative covariances induced by standardizing
$\bld{X}$. Such negative covariances should diminish as the sample
size increases, so we expect the ratio to generally be no smaller
than one, indicating that the naive variance does underestimate uncertainty.
Finally, the largest values of the ratio would occur with large positive
correlations in corresponding elements of $\mathit{\tilde{\bolds{\Sigma}}}$
and $\operatorname{Cov}(\bld{W})$, which is to be expected when both $\bld{g}$
and $\bld{X}$ show large-scale variation.

Figure~\ref{fig:prec-GLS}(c) supports these heuristic results, showing
the average ratio of variances in simulations, where the simulations
are conducted as in Section~\ref{sub:prec-ques2}. The ratio is close
to one when either of the spatial terms has fine-scale variability
and far from one when both have large-scale behavior. This result
is similar to that of \citet{Biva1980} for inference about a correlation
coefficient and to (\niekcite{JohnDiNa1997}, page 178) under serial autocorrelation
in a regression setting. As expected, when the proportion of residual
variability is smaller (moving from the bottom left to bottom right
panels), the expected ratio gets closer to one. This indicates that
when nonspatial variation dominates the residual and the spatial
structure in the residual or exposure is not too large in scale, the
naive variance estimator may be reasonable. A lack of large-scale
residual structure might result from having accounted for large-scale
variation in attempting to reduce spatial confounding bias. Results
with gridded locations show ratios slightly closer to one, and with
clustered locations, ratios further from one. Note that the uncertainty
estimate in any given naive analysis may be larger than when fitting
a spatial model because the more sophisticated model both corrects
the variance estimate, which increases the estimated uncertainty,
and uses a more efficient estimator, which decreases the fundamental
uncertainty.

Simple simulations with spatial ranges and sampling designs specific
to an analysis could be easily carried out for further guidance in
a given setting, allowing one to assess whether ignoring the spatial
structure has substantial impact on uncertainty estimation. Accounting
for small-scale spatial correlation requires estimation of the spatial
structure and is often computationally burdensome, so an assumption
of independence can have an important practical benefit. Of course
in some analyses, any underestimation of variability may be cause
for concern, in which case use of the naive variance estimator would
not be tenable.

\section{Case Study: Birthweight and\break Air Pollution}\label
{sec:Case-study:-Massachusetts}

Chronic health effects of ambient air pollution in developed countries
involve small relative risks, but are of considerable public health
importance because of widespread exposure. Epidemiologic studies attempt
to estimate a small effect from data with high levels of variability
and stronger effects from other covariates, including potential confounders
such as socioeconomic status, so spatial confounding bias is of critical
concern.

I reanalyze data on the association between ambient air pollution
(estimates of black carbon, a component of particulate matter) and
birthweight in eastern Massachusetts \  (\niekcite{Zekaetal2008}; \niekcite{Grypetal2009}).
These analyses found significant negative effects of traffic proxy
variables and black carbon, respectively, on birthweight. \citet{Grypetal2009}
used several methods to try to account for effects of measurement
error in the predicted black carbon concentrations, which are based
on a regression model that accounts for spatial and temporal structure
and key covariates.

I follow these analyses in using an extensive set of covariates to
try to account for potential confounding. I~use smooth terms for mother's
age, gestational age and mother's cigarette use, to account for nonlinearities,
a linear term for census tract income, and categorical variables for
the following: presence of a health condition of the mother, previous
preterm birth, previous large birth, sex of baby, year of birth, index
of prenatal care and maternal education. The exposure of interest
is the estimated nine-month average black carbon concentration at the
geocoded address of the mother, based on a black carbon prediction
model \citep{Grypetal2007}. Following \citet{Grypetal2009},
for simplicity, I exclude the 13,347 observations with any missing
covariate values, giving 205,713 births.

\begin{figure*}

\includegraphics{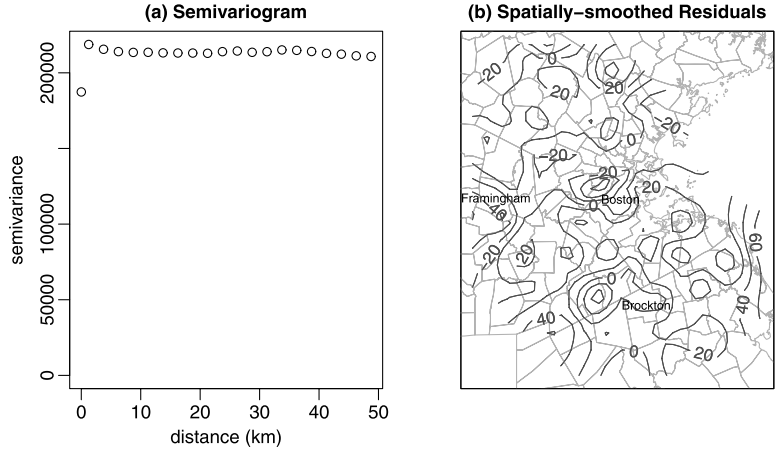}

\caption{\textup{(a)} Semivariogram of full model residuals, with the first
point representing
births to mothers living at the same location. \textup{(b)} Spatial smooth
of residuals with town boundaries in grey. The spatial smooth, with
129 e.d.f. chosen by GCV, is highly significant.}\label{fig:resids}
\end{figure*}

\begin{figure*}[b]

\includegraphics{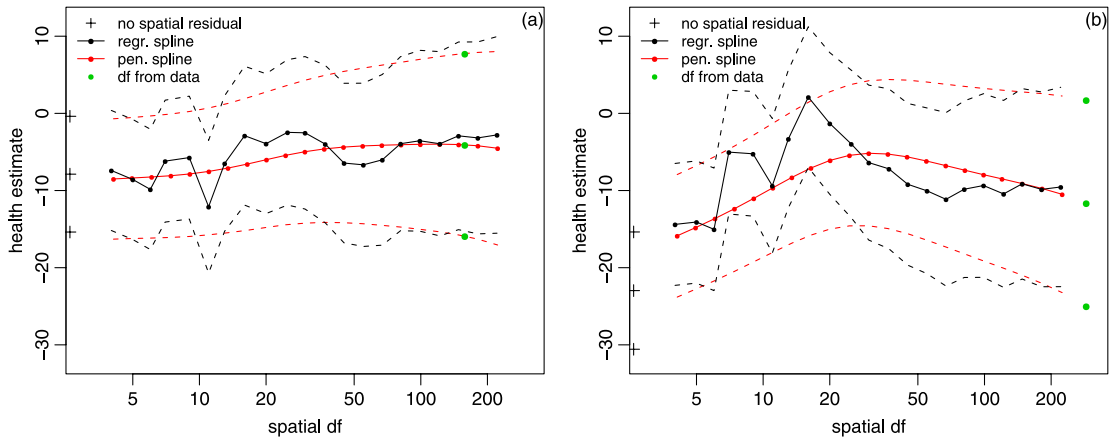}

\caption{For the model with the full set of covariates \textup{(a)} and the reduced
set of covariates \textup{(b)}, black carbon effect estimates and 95\% confidence
intervals based on different specifications for the spatial term in
an additive model: black pluses indicate the model with no spatial
term and green dots with the e.d.f. chosen by GCV, while black (regression
spline) and red (penalized spline) dots indicate results when fixing
the degrees of freedom at a set of discrete values. The lines through
the points and corresponding dashed lines are taken by connecting
the effect estimate and confidence interval bounds for the discrete set.}\label{fig:fullResults}
\end{figure*}

In \citet{Grypetal2009} we found no evidence of residual spatial
correlation based on a spatial semivariogram. Further analysis here
indicates that there is significant residual spatial variation but
that nonspatial variation overwhelms the magnitude of this variation.
Figure~\ref{fig:resids}(a) is a semivariogram showing no evidence of
spatial structure, while a spatial smooth of model residuals
[Figure~\ref{fig:resids}(b)]
indicates clear spatial structure. While individual nonspatial variability
among babies swamps the spatial variation (hence the flat semivariogram),
it is large relative to the estimated pollution effect (note the surface
values in the range of $-$40 to 40, for comparison with effect estimates
in Figure~\ref{fig:fullResults}). Thus, if the residual spatial variation
is caused by spatially varying confounders, it could bias estimation
of the pollution effect.

%
%

%
%

To include a spatial term in models of birthweight, I consider a regression
spline, an unpenalized approach, and a penalized spline, both with
e.d.f. chosen in advance (see Section~\ref{sub:Bias-variance}), as well
as a penalized spline with data-driven smoothing parameter estimation
based on GCV, all implemented in gam() in R, using the thin plate
spline basis. Note that the thin plate regression spline approach
implemented in gam() should minimize sensitivity to knot placement
\citep{Wood2006}.

I first add a spatial term to the model with the full set of covariates
to assess whether some of the estimated effect may be biased by spatial
confounding. Figure~\ref{fig:fullResults}(a) shows how the estimated
effect of black carbon varies with the e.d.f. and the spatial smoothing
approach. The estimate attenuates somewhat as more e.d.f. are used to
account for the spatial structure. For the penalized spline, as more
than about 10 e.d.f. are used, the upper confidence limit exceeds zero,
and for larger e.d.f., the upper limit increases further. GCV chooses
157 e.d.f., indicating fairly small-scale spatial structure in the data.
For context note that with 129 e.d.f. in Figure~\ref{fig:resids}(b) we
see spatial features at the scale of individual towns. While the regression
spline approach implemented here avoids having to choose the knots,
the empirical results are still very sensitive to e.d.f., in contrast
to the stability of the penalized spline solution as the e.d.f. varies.
For both penalized and regression splines, there is a clear bias-variance
tradeoff, with increasing variance as the number of e.d.f. increases.
However, for this problem with a very large sample size, the confidence
intervals do not increase drastically, nor is there much difference
in the uncertainty between the regression and penalized spline approaches.
The spatial confounding assessment suggests that while we have somewhat
reduced confidence in the black carbon effect, the effect estimate
is reasonably stable even when using a spatial term with a large number
of degrees of freedom.

Next I consider what might have happened if most of the covariates
(particularly the ones related to socioeconomic status) were not measured,
potentially inducing serious confounding. Figure~\ref{fig:fullResults}(b)
indicates that without any spatial term in the model, the effect estimate
is $-$23.0 with a 95\% confidence interval of ($-$26.8, $-$19.2), indicating
a much more substantial effect of black carbon than the fully adjusted
model. As soon as one accounts for spatial structure, even with a
small number of e.d.f., the estimate attenuates, approaching the fully
adjusted estimate, with the upper confidence limit rising above zero.
The reduced model appears to suffer from serious confounding, with
the estimated pollution effect apparently driven by large-scale association
of pollution and birthweight. The spatial analysis is able to account
for much of this apparent confounding, substituting for a rich set
of covariates.

Ideally one would fit a model that accounts for fine-scale spatial
structure to improve one's confidence in the uncertainty estimation.
However, with 205,713 observations, this is a computational challenge
that I do not take up here. Given the results in Section~\ref{sub:prec-ques3}
that indicate that large-scale structure causes most of the variance
underestimation, one can hope that the uncertainty at the larger values
for the spatial e.d.f. in Figure~\ref{fig:fullResults} may reasonably
approximate the true uncertainty.

\section{Discussion}

Considerations of scale are critical in spatial regression problems.
Standard spatial regression \mbox{models}, which use spatial random effects,
kriging specifications or a penalized spline to represent the spatial
structure, are penalized models with inherent bias-variance tradeoffs
in estimating the smooth function. Under unmeasured spatial confounding,
the bias carries over into estimating the coefficient for the exposure
of interest, but the degree of bias depends on the spatial scales
involved. Inclusion of a spatial residual term accounts for spatial
correlation in the sense of reducing bias from unmeasured spatial
confounders only when there is unconfounded variability in the exposure
at a scale smaller than the scale of the confounding. If the variation
in exposure is solely at large scales, there is little opportunity
to reduce spatial confounding bias, but with a component of small-scale
exposure variability, large-scale spatial confounding bias can be\break
reduced substantially. Accounting for large-scale\break residual correlation
is also important for improving precision of regression estimators
and for correctly estimating uncertainty. In contrast, when residual
correlation occurs at small scales, there is little opportunity for
reducing spatial confounding bias at those scales or improving regression
estimator precision. However, under the assumption of no small-scale
confounding, fitting such residual structure can reduce bias from
larger scale confounding by causing undersmoothing with respect to
the large-scale structure. While the results here are limited to the
simple setting of linear regression/additive models with a single
covariate and single unmeasured confounder, I expect that the qualitative
results and principles hold in more complicated settings, with no
reason to believe that the bias results would improve in more complicated
models.

Sensitivity analyses that show the bias-variance tradeoff as a function
of the scale at which\break the spatial residual structure is modeled\break (\niekcite{Pengetal2006}; \niekcite{Zegeetal2007})
offer one approach that helps to frame the issue of bias in the context
of the spatial scales involved. In choosing a spline formulation to
carry out such an analysis, while a regression spline has an appealing
interpretation and in theory result in less bias in estimating the
effect of interest, a penalized spline with a fixed effective degrees
of freedom may give more stable results. Of course the sensitivity
analysis approach does not answer the question of how to get a single
estimate of the effect of interest. One might also consider an approach
similar to that of \citet{Beeletal2007} and explicitly decompose
the exposure into multiple scales, including exposure at each scale
as a separate covariate and focus causal interpretation on the effect
estimates for the smaller scales (e.g., \niekcite{Janeetal2007}).
\citet{LuZege2007} use matching estimators for each pair of observations
and assess how effect estimates vary with spatial lag between the
pairs to assess sensitivity. Note that estimating equation approaches
are not capable of reducing bias from unmeasured spatial confounding
because the marginal variance is assumed to be unrelated to the exposure
and variation is not attributed to a spatial term.

From the econometric perspective, spatial\break confounding bias might be
seen as a type of endogeneity bias, with exposure the endogenous variable
and the unconfounded component of exposure, or some proxy for it,
an exogenous variable. Since the unconfounded component is not measured
directly, some sort of scale decomposition appears necessary. Standard
endogenous variable techniques such as two-stage least squares and
instrumental variable methods \citep{JohnDiNa1997} do not appear
directly useful but do share commonalities with approaches mentioned
above.

Others have noted the identifiability problems in spatial models,
with sensitivity of effect estimates to inclusion of a spatial residual
term when the\break  covariates \ vary \ spatially \ (\niekcite{BresClay1993};
\niekcite{Clayetal1993};\break \niekcite{Burdetal2005}; \niekcite{Laws2006}, page
187;\break
\niekcite{Auguetal2007}; \niekcite{Wake2007}). A different methodologic
perspective than that presented here has been taken by \citet{Reicetal2006}
and \citet{Housetal2006}, who estimate the effect of exposure,
$\bld{X}$, by forcing the spatial residual to be orthogonal to $\bld{X}$,
attributing as much variability as possible to $\bld{X}$. This approach
makes a very strong assumption of no confounding to avoid overadjustment
bias from accidentally accounting for some of the effect of the covariate
in the residual. Note that the residuals and covariates are not orthogonal
under GLS estimation (Schabenberger and Gotway, \citeyear{SchaGotw2005}, page 349). \citet{GustGree2006}
confront a similar problem of modeling systematic residual confounding
in a context with identifiability problems, finding that imposing
structure through a prior distribution in a nonidentified model can
help account for a portion of the confounding, improving bias and
precision of estimators.

Note that measurement error in the exposure is of critical concern,
because reducing bias relies on estimating variability in exposure
at scales smaller than the confounding. In many contexts, measurement
error becomes an increasing concern at small scales\break because of limitations
in measurement resources. In contrast, large-scale exposure variation
may be well estimated using spatial smoothing and regression models,
thereby inducing Berkson-type error\break through what is effectively regression
calibration\break \citep{Grypetal2009}. To the extent to which accounting
for bias forces one to rely on exposure estimates more likely contaminated
by classical measurement error, one may find oneself reducing bias
from confounding only to increase it from measurement error. To the
extent small-scale variation is affected by Berkson error, one would
increase variance but not incur bias by relying on the small-scale
variation.

Finally note that in many settings one has aggregated exposure and
outcome data, so one has limited ability to identify effects of exposure
based on fine-scale variation because the aggregation eliminates the
fine-scale variation (e.g., Janes, Dominici and Zeger, \citeyear{Janeetal2007}). This suggests
that accounting for spatial confounding with areal data, for
which\break
researchers often use standard conditional\break auto-regressive models,
is likely to be ineffective\break when aggregating over large areal units,
which is consistent with the bias seen in \citet{Rich2003}. In work
concurrent with that presented here, \citet{HodgReic2010} have investigated
bias in the areal setting under a variety of perspectives on the spatial
random effects, also making the case for the approach taken in \citet{Reicetal2006}.

\begin{appendix}

\section*{\texorpdfstring{Appendix: Proof of Lemma~\protect\ref{LEM}}{Appendix: Proof of Lemma 3.1}}\label{append}

From the definition of the GLS estimator, we have
\begin{eqnarray*}
\operatorname{Var}(\hat{\beta}_{\smttt{GLS}})
&=&[\mathit{\bolds{\mathcal{X}}}^{T}\mathit{\bolds{\Sigma}}^{-1}\mathit{\bolds{\mathcal{X}}}]_{2,2}^{-1}
\\
&=&\frac{\bld{1}^{T}\mathit{\bolds{\Sigma}}^{-1}\bld{1}}{\bld{1}^{T}\mathit{\bolds{\Sigma}}^{-1}
\bld{1}\bld{X}^{T}\mathit{\bolds{\Sigma}}^{-1}\bld{X}
-\bld{X}^{T}\mathit{\bolds{\Sigma}}^{-1}\bld{1}\bld{1}^{T}\mathit{\bolds{\Sigma}}^{-1}\bld{X}}.
\end{eqnarray*}
Using the definitions of $\mathit{\tilde{\bolds{\Sigma}}}$ and $p_{g}$,
and taking the reciprocal, we have
\begin{eqnarray*}
\operatorname{Prec}(\hat{\beta}_{\smttt{GLS}})
&=&\frac{1}{\sigma_{g}^{2}+\tau^{2}} \biggl(\bld{X}^{T}\mathit{\tilde{\bolds{\Sigma}}}{}^{-1}\bld{X}
\\
&&\hspace*{42pt}{}-\frac{\bld{X}^{T}\mathit{\tilde{\bolds{\Sigma}}}{}^{-1}
\bld{1}\bld{1}^{T}\mathit{\tilde{\bolds{\Sigma}}}{}^{-1}\bld{X}}{\bld{1}^{T}\mathit{\tilde{\bolds{\Sigma}}}{}^{-1}\bld{1}} \biggr).
\end{eqnarray*}
Conclude by taking the expectation with respect to the sampling distribution
of $\bld{X}$, using the expectation of a quadratic form, and rearranging
the matrices inside the second trace to give a scalar:
\begin{eqnarray*}
&&\hspace*{-5pt}\mathrm{E}_{X} (\operatorname{Prec}(\hat{\beta}_{\smttt{GLS}}) )
\\
&&\hspace*{-5pt}\quad  =
\frac{\sigma_{x}^{2}}{\sigma_{g}^{2}+\tau^{2}} \biggl(\operatorname{tr}(\mathit{\tilde{\bolds{\Sigma}}}{}^{-1}\mathit{\bld{R}}(\theta_{x}))
\\
&&\hspace*{-5pt}\qquad\hspace*{43pt}{}
-\frac{\operatorname{tr}(\mathit{\tilde{\bolds{\Sigma}}}{}^{-1}\bld{1}\bld{1}^{T}\mathit{\tilde
{\bolds{\Sigma}}}{}^{-1}\mathit{\bld{R}}(\theta_{x}))}{\bld{1}^{T}\mathit{\tilde{\bolds{\Sigma}}}{}^{-1}\bld{1}}  \\
&&\hspace*{-5pt}\qquad\hspace*{43pt}{}+  \mu_{x}^{2}\bld{1}^{T}\mathit{\tilde{\bolds{\Sigma}}}{}^{-1}\bld{1}
-\frac{\mu_{x}^{2}\bld{1}^{T}\mathit{\tilde{\bolds{\Sigma}}}{}^{-1}\bld{1}\bld{1}^{T}
\mathit{\tilde{\bolds{\Sigma}}}{}^{-1}\bld{1}}{\bld{1}^{T}\mathit{\tilde{\bolds{\Sigma}}}{}^{-1}\bld{1}} \biggr)\\
&&\hspace*{-5pt}\quad =  \frac{\sigma_{x}^{2}}{\sigma_{g}^{2}+\tau^{2}} \biggl(\operatorname{tr}(\mathit{\tilde{\bolds{\Sigma}}}{}^{-1}\mathit{\bld{R}(\theta
_{x})})-\frac{\bld{1}^{T}\mathit{\tilde{\bolds{\Sigma}}}{}^{-1}\mathit{\bld
{R}(\theta_{x})}\mathit{\tilde{\bolds{\Sigma}}}{}^{-1}\bld{1}}{\bld
{1}^{T}\mathit{\tilde{\bolds{\Sigma}}}{}^{-1}\bld{1}} \biggr).
\end{eqnarray*}
\end{appendix}

\section*{Acknowledgments}
The author thanks Louise Ryan and Francesca Dominici for feedback
and encouragement, Andy\break Houseman, Eric Tchetgen and Brent Coull for
comments, Ben Armstrong and John Rice for thought-provoking discussions,
Joel Schwartz for access to the birthweight data, Alexandros Gryparis
and Steve Melly for assistance with the birthweight data, and Brent
Coull for funding through NIEHS R01 Grant ES01244. This work was also
funded by NIEHS Center Grant ES000002 and NCI P01 Grant CA134294-01.

\end{document}